\documentclass[10pt,conference]{IEEEtran}

\usepackage{cite}
\usepackage{amsmath,amssymb,amsfonts}
\usepackage{algorithmic}
\usepackage{graphicx}
\usepackage{textcomp}
\usepackage{xcolor}
\usepackage{url}

\usepackage{etoolbox} 

\begin{document}

\title{What is unethical about software?\\User perceptions in the Netherlands}

\author{
\IEEEauthorblockN{
Yagil Elias\textsuperscript{1}, 
Tom P. Humbert\textsuperscript{1}, 
Lauren Olson\textsuperscript{1}, 
Emitzá Guzmán\textsuperscript{1}}
\IEEEauthorblockA{\textsuperscript{1}VU Amsterdam, Amsterdam, The Netherlands\\
Emails: yagilelias@gmail.com, t.humbert@vu.nl, l.a.olson@vu.nl, e.guzmanortega@vu.nl}
}

\maketitle

\begin{abstract}
Software has the potential to improve lives.
Yet, unethical and uninformed software practices are at the root of an increasing number of ethical concerns. Despite its pervasiveness, few research has analyzed end-users perspectives on the ethical issues of the software they use. We address this gap, and investigate end-user's ethical concerns in software through 19 semi-structured interviews with residents of the Netherlands.
We ask a diverse group of users about their ethical concerns when using everyday software applications. 
We investigate the underlying reasons for their concerns and what solutions they propose to eliminate them. We find that our participants actively worry about privacy, transparency, manipulation, safety and inappropriate content; with privacy and manipulation often being at the center of their worries.
Our participants demand software solutions to improve information clarity in applications and provide more control over the user experience.
They further expect larger systematic changes within software practices and government regulation.

\end{abstract}


\section{Introduction}
Software has become an integral part of modern life, embedding itself into almost every aspect of our daily routines. Software systems are ubiquitous and unavoidable, from social interactions and personal finance to healthcare and education. As technology advances and integrates deeper into our social fabric, ethical concerns regarding software are also on the rise. 
These \textit{ethical concerns} often include issues related to privacy, manipulation, and safety, among others.

Recent surveys highlight the magnitude of public anxiety regarding the ethical implications of emerging technologies. A Pew survey~\cite{Faverio_Tyson_2023} found that most people in the US are more concerned than excited about the impact of artificial intelligence. Similarly, a global survey~\cite{IPSOS_2023} reports that over 85\% of people are worried about the influence of online disinformation, and 87\% believe it has already harmed their country's politics. Parents also express a significant level of concern over their children's interactions with technology, with 76\% of parents deeming it critical to manage their children's smartphone use~\cite{Auxier_Anderson_Perrin_Turner_2020}. 
Another recent Pew survey\cite{pewresearchAmericansView} found that people in the US believe they have little to no control over how companies (73\%) or the government (79\%) use their personal data. The same survey finds that 56\% frequently skip reading online privacy policies as 61\% are skeptical if their privacy choices have any real impact.
These findings illustrate the urgent need to address a broadening spectrum of ethical concerns, as novel technologies add new worries faster than they can be solved.   

Until now, research has mainly addressed these ethical concerns in a compartmentalized manner, asking end-users about specific issues such as privacy~\cite{Coghlan2021, Neteler2024} or focusing on individual technologies~\cite{chang2024, Shahriar2023, Ahmed2024, Akgun2022, Borenstein2021, weidinger2021, raji2020}. Nevertheless, in practice, ethical concerns are deeply intertwined---privacy, manipulation, and safety concerns, for example, frequently overlap and interact with each other. 
Research that analyzes end-users ethical concerns about software in a more interspersed manner is sparse, and this existing body of work mainly uses data science techniques on social media posts for its analysis~\cite{tushev2020digital, besmer2020investigating, reddit_privacy, khalid2014mobile, shams2020society, obie2021first, schwartz2012overview, olson2023}. However, not everyone shares their experiences or worries online~\cite{tizard2022voice}, which leads to an incomplete understanding of how these concerns vary across diverse demographic and cultural contexts.

This study aims to explore the breadth and interconnectedness of ethical concerns in software held by a diverse group of residents of the Netherlands.
We intentionally cast a wide net, not restricting answers to any software domain, nor selecting a specific type of user.
We believe that the common themes that arise will paint a `big picture'; i.e., a map of ethical concerns pointing to areas to be examined in detail, of use to policy makers and software practitioners. Using an interpretive approach, we analyze the experiences of 19 participants in the context of their everyday software usage. The analysis involves deductively coding different types of ethical concerns and identifying recurring themes and underlying factors that influence these concerns. We also examine participants' views on potential changes to software that could alleviate their concerns, providing insights into how different ethical issues are interrelated, and suggesting directions for future software design and policy improvements.

This work contributes (1) a locally and temporally relevant overview of ethical concerns according to end-users, (2) knowledge about the interplay of concerns and how they relate to perceived issues, and (3) suggestions for solutions from the perspectives of end-users.
To encourage further research, we share a replication package that contains the interview protocol, the consent form, and annotation guidelines at \url{https://doi.org/10.6084/m9.figshare.27193323}.

\section{Related Work}

\subsection{Ethical responsibility in software practices}

A history of misalignment of human values and software practices' values has led to a dearth of research on end-user software ethics.
Practitioners often see software as devoid of values~\cite{whittle2019is}.
Although, there have been many positive changes in the software industry with many companies introducing progressive codes of ethics.
But, internal ethics do not necessarily lead to ethical software~\cite{whittle2019is, mcnamara2018}.
Practitioners still implement software according to personal values because there is no common understanding or method to elicit ethical requirements~\cite{alidoosti2022incorporating}.
Despite not having standard processes, we find a growing body of formal methods to integrate ethics in software practices, for example, by introducing formal methods to capture software ethics~\cite{drechsler2019code, paech2020how, Halme2024, Chivukula2024} and to implement ethical software~\cite{autili2019asoftware, Gogoll2021}.

Software ethics education is necessary and will lead to greater awareness of ethical responsibilities and encourage practitioners to evaluate how their work influences society~\cite{kumar2019integrating, garcia2024}.
Tahaei et al.~\cite{Tahaei2021} show that individual advocates for issues such as security or privacy help to steer their co-workers towards more ethical choices.
Widder et al.~\cite{Widder2023} find that software practitioners have ethical concerns about military, privacy, advertising, surveillance and more, sometimes even questioning the reason why their company exists.

The operationalization of ethics into software practices and education requires an understanding of current ethical issues.
A recent wave of research has delved into specific areas of concern, such as
user privacy and data security~\cite{Coghlan2021,li2022narratives,nema2022analyzing, Neteler2024},
algorithmic discrimination~\cite{zliobaite-2017},
online scamming/abuse~\cite{Amirkhani2024, Janeiro2024},
and manipulation through design choices~\cite{lacey2019cuteness}.
We also find research on ethical concerns about specific technologies such as artificial intelligence (AI), in general~\cite{peters2020, chang2024, Shahriar2023, Ahmed2024}, in education~\cite{Akgun2022, Borenstein2021}, large language models (LLM)~\cite{weidinger2021}, and facial recognition~\cite{raji2020}.

Yet, there is little research that connects the dots between ethical concerns~\cite{johnson2021towards}.
In our work, we establish an understanding of how software affects people.
Mapping user's worries is necessary to understand where software changes can have the greatest impact.

\subsection{The voice of the end-user}
User feedback is essential for building useful and relevant software~\cite{lin2000relationship}. This feedback can be obtained directly from surveys or interviews, or indirectly through data mining, among other methods.  
Previous studies in software engineering have  examined users' ethical concerns through mining App Stores or social media platforms. For example, studies have focused on mining data about discrimination~\cite{tushev2020digital}, privacy~\cite{besmer2020investigating, reddit_privacy, khalid2014mobile}, and human values violations using the Schwartz theory~\cite{shams2020society, obie2021first, schwartz2012overview}. Other work has analyzed ethical concerns expressed in app reviews~\cite{Tjikhoeri2024} and Reddit~\cite{olson2023}. Similarly, previous work has studied the interrelated nature of ethical concerns expressed in app reviews through pattern mining~\cite{karaccam2024uncovering}.

To our best knowledge, there is no work that has used interview studies to explore the full extent of users' ethical concerns in software. 
Nevertheless, there are interview studies that examine users' ethical concerns surrounding specific technology. For example, Mahdavi Goloujeh et al.~\cite{MahdaviGoloujeh2024} interviewed users on their views on ethics and prompt enginnering, while Freeman et al.~\cite{Freeman2024} used interviews to investigate users views on ethics in virtual reality spaces and found new worries surrounding privacy and cyberbullying. Amirkhani et al.~\cite{Amirkhani2024} interviewed victims of online dating abuse and highlight the importance of studying specific concerns in a local context. We take this last recommendation into consideration when designing our study. Our work presents an exploration of the ethical concerns about software in the Netherlands, the top ranking European country in terms of privacy concerns~\cite{ESS10}.
New developments in the software domain often have unforeseen effects on social dynamics, requiring a continuous conversation with end-users to uncover new, unknown ethical issues. This study hopefully opens this conversation.

\begin{table*}[t]
\renewcommand{\arraystretch}{1.3}
\caption{Ethical concern taxonomy used to create, guide and evaluate the interviews. (*) Inquired about specifically. (**) Added to taxonomy during open coding.}

\begin{center}
\begin{tabular}{|p{0.14\linewidth}|p{0.81\linewidth}|}
\hline
\textbf{Ethical concern} & \textbf{Definitions} \\

\hline

Accessibility & The participant raises concerns of the application not including people with special needs or disabilities.\\ 
\hline

Accountability & The user experienced an issue when using the application or its service. The user could not find the software company responsible for   solving the issue.  \\ 
\hline

Content theft & Content from a user is stolen or used without permission from the original creator. \\ 
\hline

Cyberbullying & The platform’s community is being harmful, abusive, or unhealthy by practicing hateful communication via the application.  \\ 
\hline

Harmful advertising & The user notices the presence of deceiving, misleading, or harmful advertisements throughout the application. \\ 
\hline

Identity theft & Someone is using the identity of someone else on this application. This concern also applies to catfishing, creating fake profiles to trick and deceive other users on the platform. \\ 
\hline

Inappropriate content & The application contains content other than advertisements that are disturbing to certain groups of people. \\ 
\hline

Misinformation & False information is spread through this application.\\
\hline

Scam & The user has been scammed or came into contact with a scammer through the application. This concern can occur through the application only or its services. A scammer deceives another to gain something, usually money or goods. \\ 
\hline

Privacy & The users' identity and data are not kept secure or used for purposes other than what the user gave consent to. This concern also includes when an account is hacked.  \\ 
\hline

Safety & The usage of this app has led to health issues or other safety risks. This concern can be about the usage of the application itself or its services. \\ 
\hline

Sustainability & The user says something about the negative impact the application has on the environment. \\ 
\hline

Transparency & The motives, risks, and implications are unclear to the user when using this application or a service the application provides.\\ 

\hline
Addiction (*) & The user mentions how they or other users are addicted to the application or describes that they use it excessively. 

\textit{E.g., ``When someone surfs the web too much, leading to decreased productivity at work and/or fewer interactions with family members."} \\

\hline
Censorship (*) & The application deliberately hides certain information, or certain users’ content or profiles are deliberately removed or demoted.

\textit{E.g., ``There is evidence that social media platforms such as Facebook and Instagram played a role in the censorship of pro-Palestinian protests."} \\

\hline
Discrimination (*) & The application user is being discriminated against by the application, its services, or its community. This concern also includes users who have an issue with having fewer functionalities available to them because they live in a different geographical area.

\textit{E.g., ``When a job application algorithm selects participants based on their ethnic background. When the software causes you to have a lower chance because of biases towards certain skin colors or certain areas."}\\ 

\hline
Manipulation (**) & The application controls, or plays upon, the user by artful, unfair, or insidious means to someone else's advantage.\\

\hline
\end{tabular}
\label{t:concern_taxonomy}
\end{center}
\end{table*}

\section{Scope and Research Questions}  
Our goal is to gather in-depth knowledge about end-users' ethical concerns when using software and potential solutions for these concerns. 
Leaning on previous work~\cite{karaccam2024uncovering}, we define \textbf{\textit{ethical concerns}} about software as end-users' worries about wrongdoing by, or through, software.
This wrongdoing can affect individuals or society as a whole.

We answer the following research questions in our study: 
\textbf{(RQ1)} \textit{What \textbf{active ethical concerns} do users have about software applications?} \textit{Active ethical concerns} are those that participants share without specifically being asked about them beforehand, i.e., those that they share in an unprimed manner.

\noindent \textbf{(RQ2)} \textit{What \textbf{latent ethical concerns} do users have about software applications?} \textit{Latent ethical concerns} are those that users describe only when explicitly being asked about them. 
Note that we call a concern latent only if a user agrees on experiencing it after being directly asked about it.

\noindent \textbf{(RQ3)} \textit{What \textbf{solutions} do users give to mitigate their ethical concerns? } In this question, we investigate which changes users view as necessary to reduce their concerns. 


\section{Method}

In this study, we analyze people's experiences with software within the context of their daily lives through an 
interpretivist lens~\cite{alharahsheh2020review}. 
We design and execute our interviews according to the the Castillo-Montoya framework~\cite{castillo-montoya-no-date} and Strandberg's ethical guidelines~\cite{strandberg-2019}. 
For this purpose, we deductively code ethical concern types in the interview transcripts, and inductively search for patterns in solutions that user's suggest to eliminate their concerns. 


\subsection{Interview design}\label{s:design}


We start our interview by explaining the purpose of the study and its structure, and giving definitions of reoccurring concepts, i.e., \textit{software application} and \textit{ethical concern}. To get a better idea of how participants use software in their daily life, we ask them how much time they spend using their phones, computers or laptops per day and to describe how they use applications on their devices in a typical day. We then follow with the main phases on \textbf{active} and \textbf{latent concerns} and their \textbf{solutions}, explained below.

%




\paragraph*{Active concerns} 
We explore active concerns by not priming participants for any specific concern.
We ask, \textit{``Were there ever times when you felt something in these applications was unethical?"}
After each answer, the interviewer requests clarifications on each newly introduced ethical concern and line of reasoning until exhaustion, i.e., when the participant cannot provide any new insights.
Each line of reasoning that has been exhausted is followed up by asking about solutions (see Section~\ref{s:solutions}).
We end this phase by asking if they have ever stopped using an app or website because they \textit{``felt something in it was unethical"}.

\paragraph*{Latent concerns}
In this phase we focus on three ethical concerns: addiction, censorship and discrimination.
We select these specific concerns as they were the most frequent concerns in previous work~\cite{olson2023} that were not mentioned in the latent phase of the four pilot interviews. 
For each ethical concern, the interviewer provides a definition (see Table~\ref{t:concern_taxonomy}), and, upon request, an example.
We ask participants if they are worried \textit{about software being addicting}, if they are worried \textit{that software applications can censor specific information}, and if they are concerned \textit{that an application, its services or community can discriminate users}.
We follow the same protocol as before to exhaust the participants' explanations of their perspectives and finalize each concern with the question about solutions (see Section~\ref{s:solutions}).
We design the interviews to be exhaustive on an individual level. This decision ensures representativeness of our sample rather than the Dutch population. Nevertheless, achieving true saturation would require an unfeasibly large sample for in-depth analysis.

\paragraph*{Solutions}\label{s:solutions}
We investigate perceived solutions of active and latent concerns with the question, \textit{``what would need to change so you would not worry anymore?"}
Every time the interviewer exhausts descriptions of the reasoning behind an ethical concern, we ask participants this question, not only to gather suggestions for software practitioners, but also to better understand underlying reasons for their worry.

We end the interview by collecting additional (optional) background information, i.e., gender, main country of influence and age. 

We conducted four pilot interviews at different stages of the design process. Based on these interviews we reformulated some of the questions. Also, we decided to limit the interview to one hour, restricting the number of latent ethical concerns we inquire to three.


\subsection{Recruitment and interview execution}
The target population are software users residing in the Netherlands. We strive for diverse gender and age representation during participant selection. Interviews are conducted between April and May 2024 in person or on MS Teams by the first author.
Participants are recruited in the Netherlands before and during that time through snowball sampling from personal contacts. 
All participants have a social distance of at least two degrees to all authors. 
We distribute a contact form to sign up, explaining the purpose of the study and the handling of personal data.
To ensure for diversity, participants are asked after the interview to ask people of a different generation and gender to participate in our study. 

Our participants are predominantly women (12/19=63\%) and Gen Z (9/19=47\%), as shown in Figure~\ref{fig:ec_frequency}.
All participants reside in the Netherlands at the time of the interviews. 
We ask participants which country they are most influenced by to ensure cultural diversity within the sample, noting that in a globalized world nationality no longer reflects cultural affiliation.
We find that 30\% (5/19) indicate the Netherlands~\raisebox{-2pt}{\includegraphics[width=0.04\linewidth]{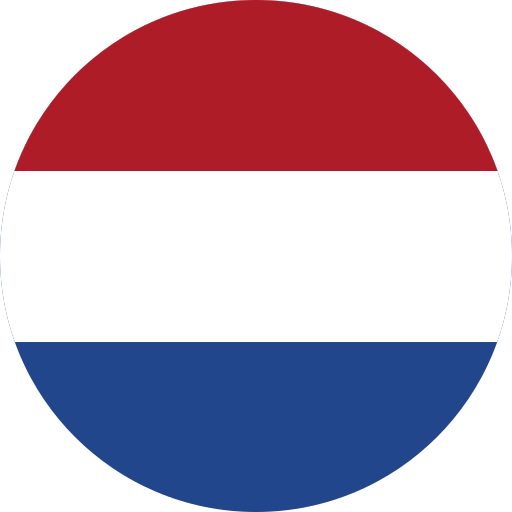}}, 
four Suriname~\raisebox{-2pt}{\includegraphics[width=0.04\linewidth]{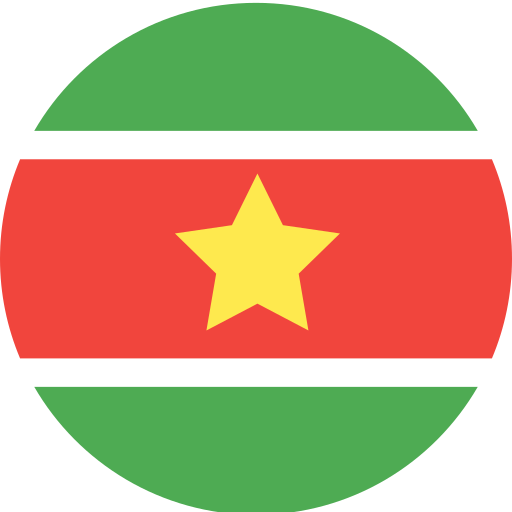}}, 
and two Kazakhstan~\raisebox{-2pt}{\includegraphics[width=0.04\linewidth]{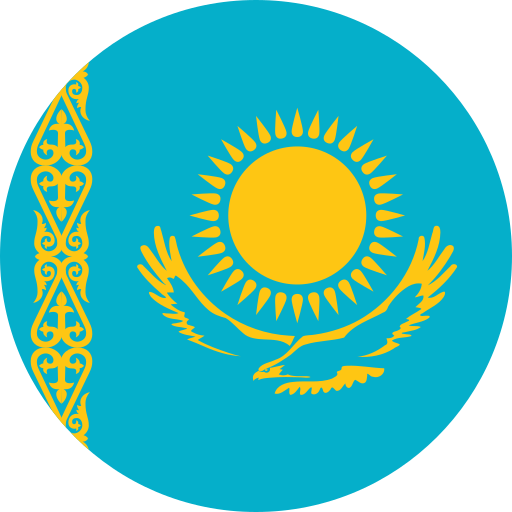}}.
The remaining participants reported being mainly influenced by Indonesia~\raisebox{-2pt}{\includegraphics[width=0.04\linewidth]{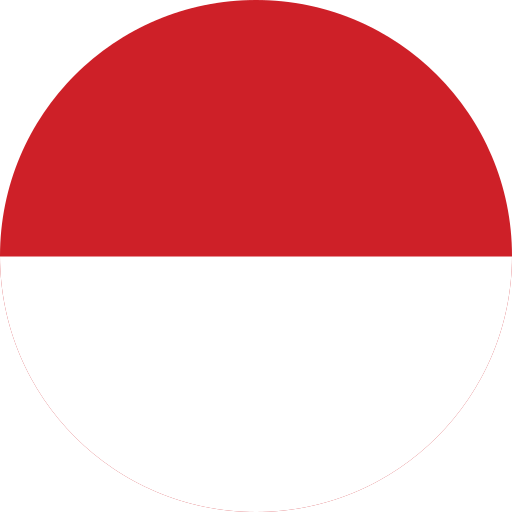}}, Azerbaijan~\raisebox{-2pt}{\includegraphics[width=0.04\linewidth]{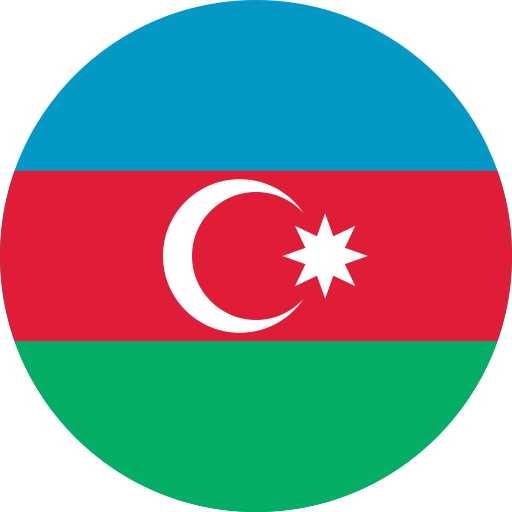}}, 
Iran~\raisebox{-2pt}{\includegraphics[width=0.04\linewidth]{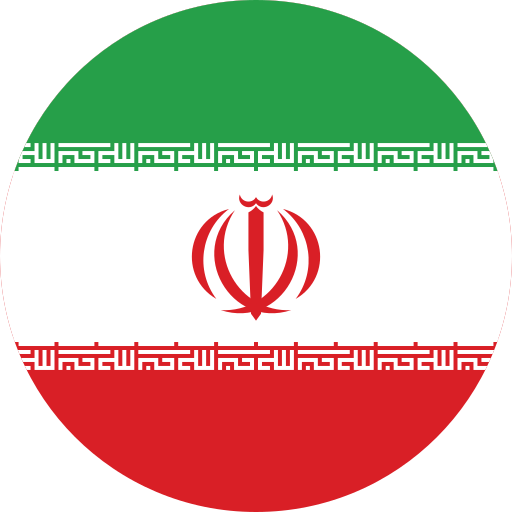}}, 
Tunisia~\raisebox{-2pt}{\includegraphics[width=0.04\linewidth]{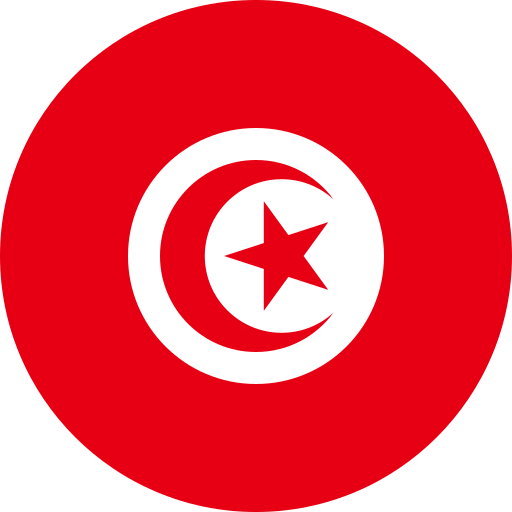}}, 
France~\raisebox{-2pt}{\includegraphics[width=0.04\linewidth]{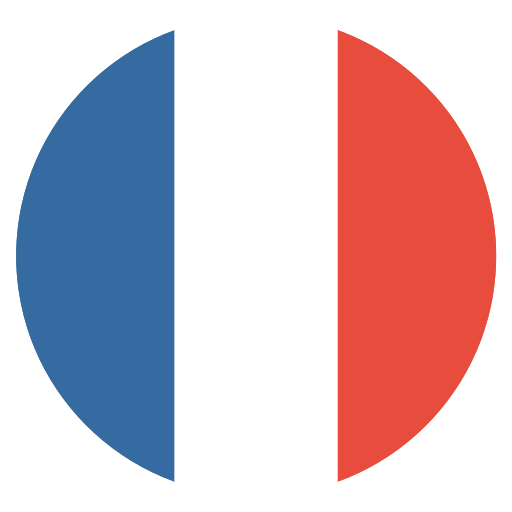}}, 
Brazil~\raisebox{-2pt}{\includegraphics[width=0.04\linewidth]{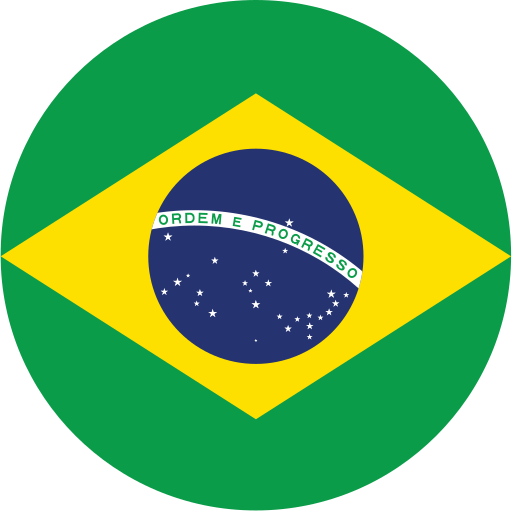}}, 
and the USA~\raisebox{-2pt}{\includegraphics[width=0.04\linewidth]{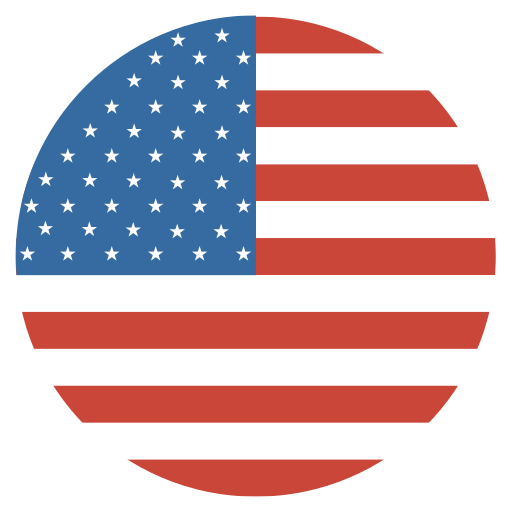}}.
Participants report a median of eight hours of software use. Everyone reports using communication or socials software on a daily basis, nine report the use of productivity software, and eight the use of entertainment software. We can roughly divide participants into two groups, of which millennials and baby boomers mostly belong to the heavy software users. 
This is due to time spent in front of a computer during work.
P8 reports having halved her time using software. P1 is the only outlier with up to 24 hours of constant software use.
Only four participants {\footnotesize[P11, P13, P16, P18]} work in the ICT sector.

\subsection{Data analysis} \label{s:data_analysis} 
We use the build-in transcription feature of MS Teams, except for the two first interviews, that have been recorded offline and have therefore been manually transcribed.
The transcripts are verbatim and only filler words are cleaned from the transcripts.
We anonymize participants through pseudonyms and report age only as the generation they belong to.
To answer RQ1 and RQ2, we deductively assign interviewee's statements with the ethical concern types from the concern taxonomy we adopted from previous work~\cite{olson2023,Tjikhoeri2024}, detailed with their definitions in Table~\ref{t:concern_taxonomy}.
Annotators can assign multiple types of concerns to participant's statements.
The two first authors executed this task independently  in four annotation rounds. 
Between each round, codes are discussed to reach perfect agreement.
The average Fleiss' kappa score before the discussions is 0.72, indicating a substantial agreement~\cite{Landis1977-bf}. 
During two larger discussions, all authors discuss ambiguous cases, and decide to add \textit{manipulation} as a new ethical concern type to the current taxonomy.
The remaining coding is done by the first author.

We then apply open and axial coding using the Atlas.ti~\cite{atlas_ti_manual} network feature to identify recurring themes of ethical concerns, and determine underlying factors.
An underlying factor is a reason an interviewee gives to explain why they hold a specific ethical concern.
We do not judge participant's perspective, even if misinformed, these underlying factors have a real influence on decisions participants will make about software.
We confine the coding process to the scope of the given ethical concern type definitions. 
For example, when privacy is detected, we look for reasons why the user is worried about their identity or data not being kept secure, or what the user did not give consent to.
An underlying factor may also be another type of ethical concern.
In the discussion, we examine how the most prevalent concerns relate to each other.

We apply open and axial coding to answer RQ3 on potential solutions for user's ethical concerns.
For this purpose, we look for patterns in participants' answers throughout all ethical concerns and group them thematically.
This means that these solutions have the potential to address multiple, intertwined ethical concerns.


\section{Findings}\label{section: findings}
\begin{figure}[!ht]
    \begin{center}
    \includegraphics[width=\linewidth]{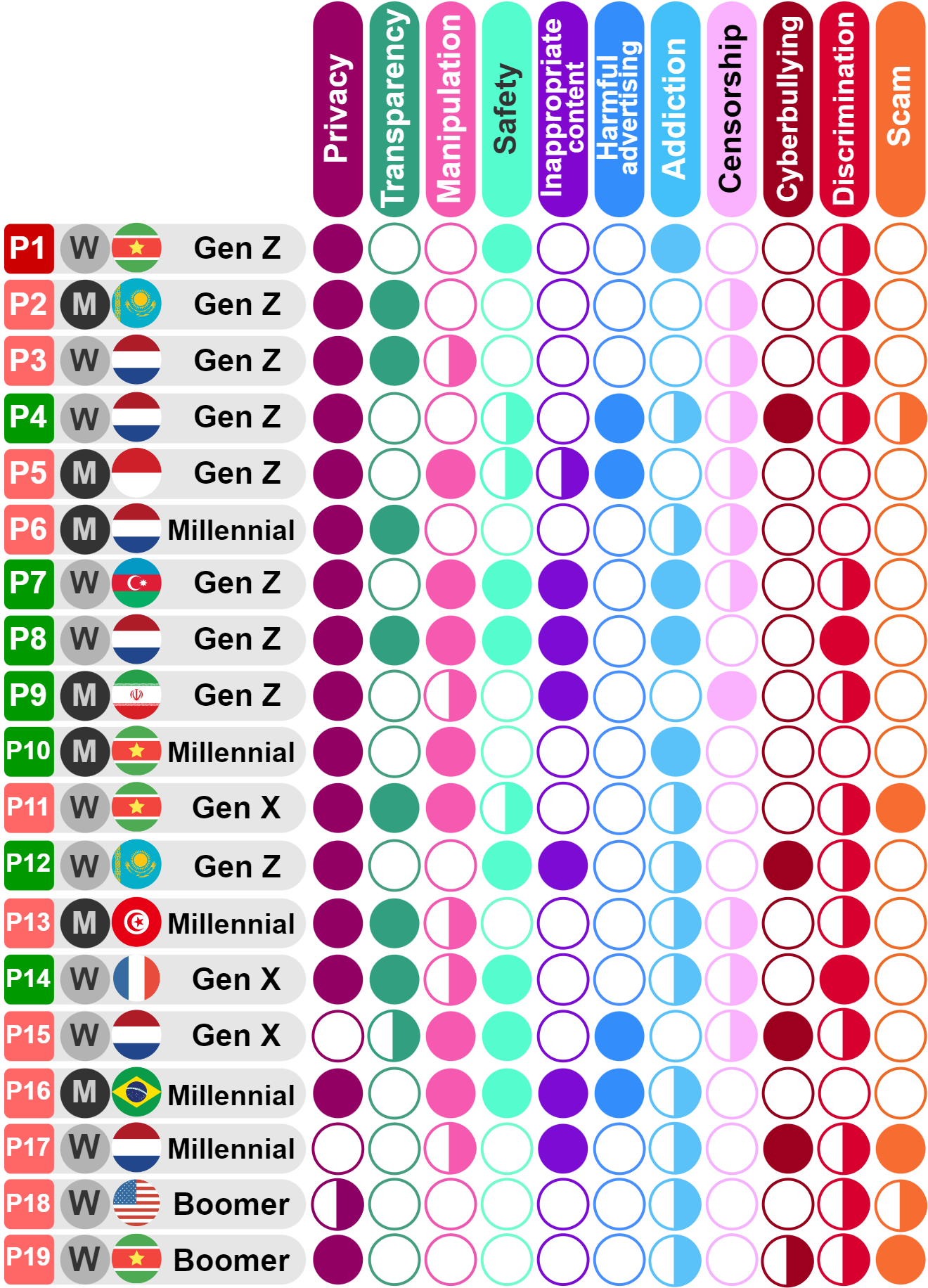}
    \caption{Participants and the ethical concerns they report in their answers. In this matrix, the rows represent the participants and their personal background, the columns represent the types of concern. 
    A full circle indicates an active ethical concern. A half circle indicates a concern that arises only when asked about the latent ethical concerns. We characterize participants by coloring heavy software users red and light users green. The letters W and M denote their gender. The flag represents the country they feel most influenced by. Participants' generation is also indicated.}
    \label{fig:ec_frequency}
    \end{center}

\end{figure}

\begin{figure}[!ht]
    \begin{center}
    \includegraphics[width=\linewidth]{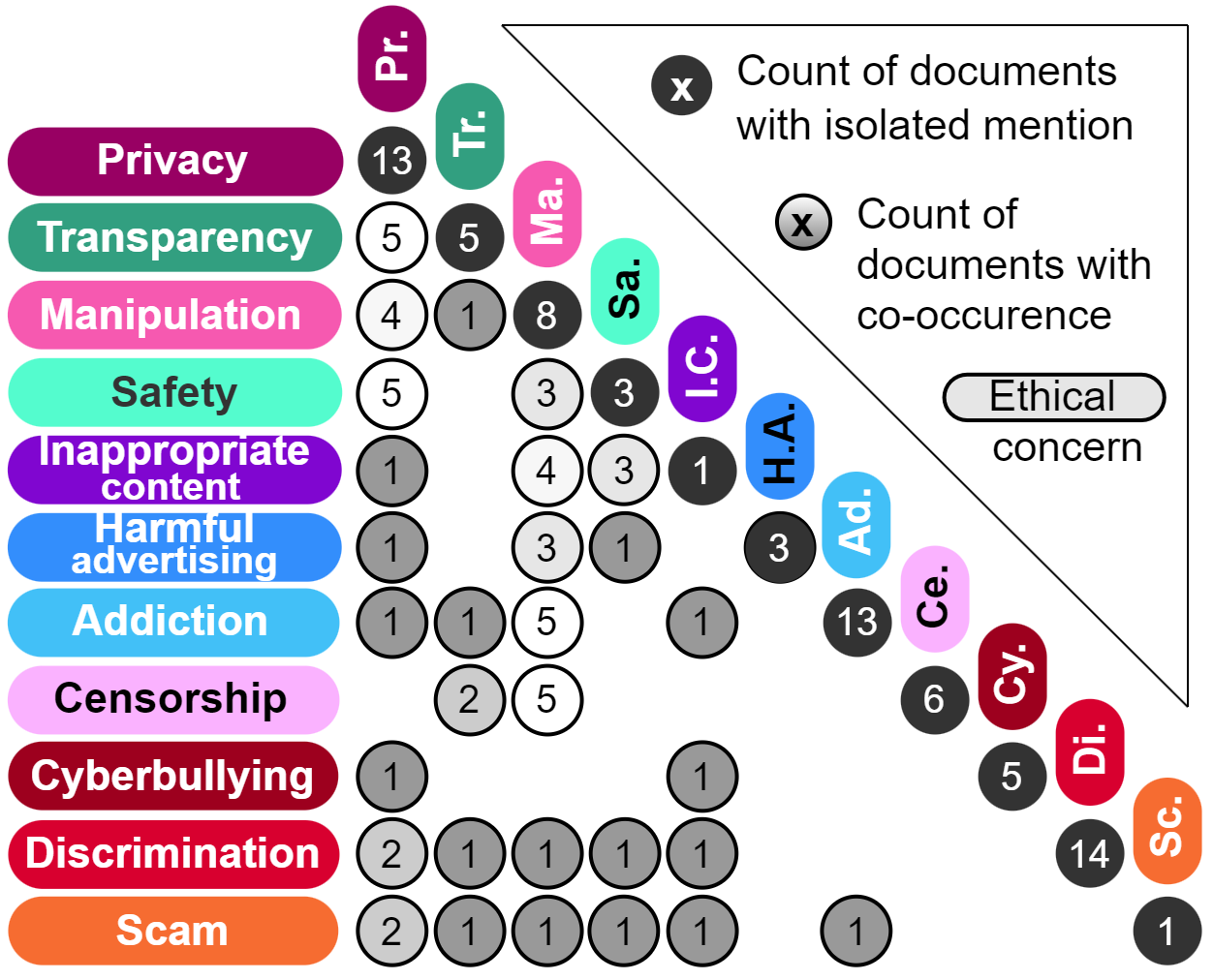}
    \caption{Co-occurrence counts of ethical concerns. The circles are colored in gray scale, where a lighter gray indicates a higher count. The count of times participants bring up an ethical concern on its own is indicated in the black circles.}
    \label{fig:ec_cooc}
    \end{center}

\end{figure}

Figure~\ref{fig:ec_frequency} illustrates the ethical concerns mentioned in each interview, highlighting if it is an active or latent concern. 
Figure~\ref{fig:ec_cooc} shows how participants' ethical concerns co-occur.
Themes around specific ethical concerns often involve additional concerns, creating an intricate weave of interrelated worries, illustrated in Figure~\ref{f:ec_map}.

\subsection{Active ethical concerns (RQ1)}
\label{sec:active}

An overwhelming majority of our participants {\footnotesize(16/19)} is actively worried about their privacy and frequently report transparency {\footnotesize(7/19)}, manipulation {\footnotesize(8/19)}, safety {\footnotesize(6/19)}, and inappropriate content {\footnotesize(6/19)} concerns.
Fourteen out of nineteen participants link personal experiences to the ethical concerns they share. 
Seven participants have stopped using an application in the past and support their decision with concerns over addiction {\footnotesize[P1, P7, P8]}, transparency {\footnotesize[P6]}, and privacy {\footnotesize[P3, P11, P14]}. 



\begin{figure*}[!ht]
    \begin{center}
    \includegraphics[width=\linewidth]{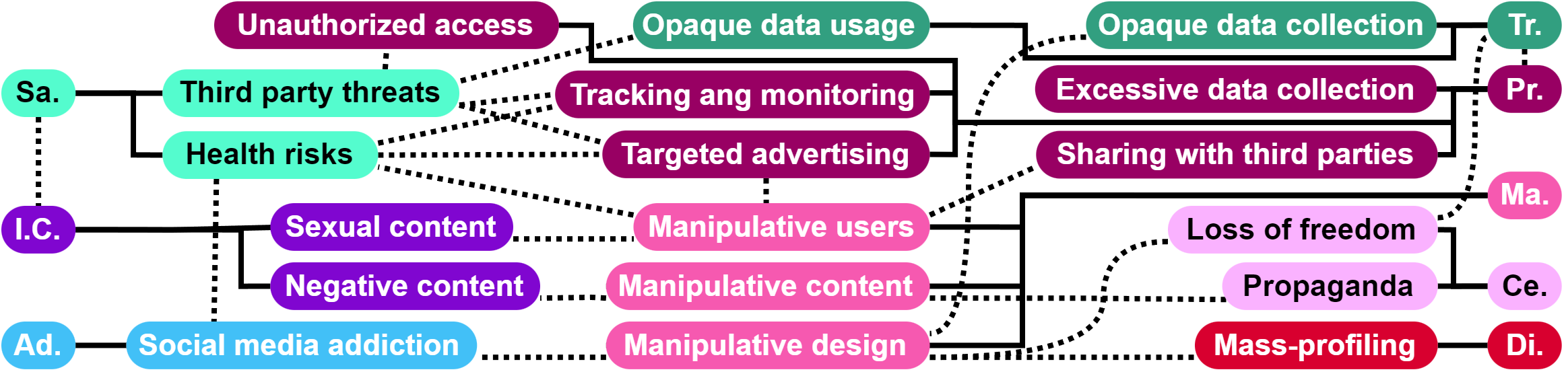}
    \caption{Types of ethical concern and their main themes. A thick line and color show how themes relate to ethical concerns. Dotted lines denote relations between concerns, inferred from our participants' reasoning.}
    \label{f:ec_map}
    \end{center}
\end{figure*}

\vspace{-2mm}
{\color{white}\tiny
\subsubsection*{Privacy}$ $\\}
\raisebox{-1.25mm}{\includegraphics[width=25mm]{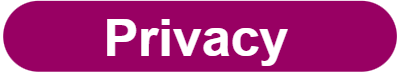}}
16 participants are actively concerned about their privacy when using software applications.
The majority of participants' remaining concerns tie back to privacy themes (cf. Figure~\ref{f:ec_map}). 


\paragraph*{Targeted advertising}
Participants are concerned about targeted advertising and perceive it as invasive and a breach of their privacy {\footnotesize[P1, P3, P7, P8, P9, P10, P13, P14, P16]}.
They often describe how ads precisely relate to some content they have viewed or information they shared in private:
\begin{quote}
 \textit{``I'll just text raincoat towards a friend, and then two minutes or hours later I go on Instagram and get an ad for a raincoat. There's no ******* way that this random ad for a raincoat is because of randomness"} - P8 GenZ Women

\end{quote}
Participants sometimes struggle to express why they perceive this practice as unethical. They might grapple to frame their worry because a fear of advertising could be perceived as irrational.
While participants may lack the technical language to explain the dangers of the technology, they are aware of potential consequences. 
We find that the omnipresence of targeted ads foments suspicions of surveillance, criminal data practices, and manipulation.
The few participants that agree targeted advertising is helpful also stress their unease about the loss of anonymity.

\paragraph*{Tracking and monitoring}
The feeling of being tracked or followed is expressed vividly {\footnotesize[P2, P3, P7, P9, P11, P12, P14, P16]}. 
This is both voiced within {\footnotesize[P7, P9, P14, P16]} and outside the context {\footnotesize[P2, P3, P11, P12, P14]} of targeted advertisements. 
There is a shared perception that software will continuously evolve towards collecting more personal data in more intrusive ways.

\paragraph*{Excessive data collection}
Five participants {\footnotesize[P1, P3, P11, P12, P14]} note that data is collected in excess.
Two participants believe these masses of data to be irrelevant to the functionality of the software application {\footnotesize[P11, P12]}.



\paragraph*{Sharing with third-parties}
Furthermore, a couple of participants express their concern towards companies sharing personal data with third-parties such as marketing, or government agencies {\footnotesize[P5, P11]}.
P5 is worried about political manipulation on social networks, mentioning the Cambridge Analytica scandal~\cite{theguardianMadeSteve}.

\paragraph*{Unauthorized access}
The same participants share their concern about the protection of their personal data from unauthorized access {\footnotesize[P5, P11]}.


The underlying factors of the privacy concerns of our participants relate to concerns of transparency and manipulation.
They are  concerned because
they feel monitored or fear being \textit{``registered somewhere" }{\footnotesize[P1, P3, P9, P13, P14, P16]},
they reject that their daily life is tracked to such a degree {\footnotesize[P7, P9, P13]}, 
they feel that devices know too much {\footnotesize[P1, P3, P10]}, are unable to be anonymous online {\footnotesize[P14]},
that their thoughts are being invaded {\footnotesize[P7, P10]}, 
and that the government surveils them {\footnotesize[P5, P12]}.
They also suspect malicious reasons for why software applications collect certain personal data {\footnotesize[P12]}, pointing for example at the danger of data being used to manipulate the democratic process {\footnotesize[P5]}.
P11 perceives a lack of transparency about who can access personal information, suspecting that some companies bypass regional laws such as GDPR {\footnotesize[P11]}.

\vspace{-2mm}
{\color{white}\tiny
\subsubsection*{Transparency}$ $\\}
\raisebox{-1.25mm}{\includegraphics[width=25mm]{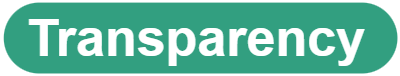}}
Seven participants express active concern about transparency in software applications.

\paragraph*{Lack of clarity on data collection}
Three Participants say it is unclear how data is collected by cookies, consent forms, and terms of usage {\footnotesize[P2, P8, P11]}.
P2 acknowledges that they could learn about this from the lengthy terms \& conditions documents of the software but;
\begin{quote}
    \textit{``It's not something that is highlighted on the first page [..], I need to read the whole consent form [to know]"} - P2 Gen Z Man
\end{quote}
They continue to explain that even if they read it, it would not be clear to them.


\paragraph*{Lack of clarity on data usage}
Six participants report that they do not know how their data is processed {\footnotesize[P3, P6, P8, P13, P14, P15]}.
They appear angered, and often cynically express a form of capitulation.
They see no other option than accepting that their personal data might be abused, feeling forced to sign terms and conditions without reading.


We find that the underlying factors of the transparency concern often relate to privacy and manipulation concerns.
Participants indicate that their transparency concerns stem from not knowing which kind of data is collected {\footnotesize[P2]}, the constant tracking of offline behavior {\footnotesize[P3]}, the suspected dishonesty about processing of personal data {\footnotesize[P6, P15]}, safety concerns {\footnotesize[P8]}, expectations of fraudulent behavior {\footnotesize[P11]}, the lack of understanding of software mechanisms {\footnotesize[P13]}, and from discriminatory practices {\footnotesize[P14]}.

\vspace{-2mm}
{\color{white}\tiny
\subsubsection*{Manipulation}$ $\\}
\raisebox{-1.25mm}{\includegraphics[width=25mm]{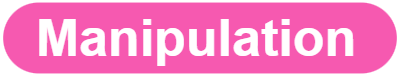}}
Seven participants actively worry about applications manipulating its users in various ways. 
Another five participants relate a latent concern to software-enabled manipulation.

\paragraph*{Manipulative design}
Seven participants believe it is unethical when applications manipulate users {\footnotesize[P7, P8, P9, P11, P14, P16, P18]}.
They describe designs to maximize user engagement {\footnotesize[P9]} either by abusing peoples emotional responses {\footnotesize[P8]}, by stimulating the dopamine system {\footnotesize[P16]}, or by constantly pushing content {\footnotesize[P11, P14]}.
P8 shares her perspective of social media:
\begin{quote}
    \textit{``Social media, apps and stuff like that, I truly believe that the way [it's] happening right now is so unethical. The way they play with emotions of people, the way they literally try to just grab your attention and keep you on an app as much as possible."} -P8 Gen Z Woman
\end{quote}
P8 and others agree that companies need to monetize the platform, but if a current practice is considered unethical it could lead to suspicions of further, hidden unethical activities.


\paragraph*{Manipulative content}
Seven participants are worried that content of applications manipulate the user {\footnotesize[P4, P5, P9, P10, P13, P15, P18]}.
Participants are worried that advertisements manipulate users' perceptions {\footnotesize[P4, P5, P10]}, leading to purchases of harmful products {\footnotesize[P4]}, influence on election outcomes {\footnotesize[P5]}, and circumvention of user awareness {\footnotesize[P10]}.


\paragraph*{Manipulative users in the application}
A couple of participants {\footnotesize[P4, P11]} mention how users can be manipulated by other users of the application, including bots that some users deploy to influence political conversations {\footnotesize[P11]}. 
P14 mentions her worry about social media influencers that make a living from convincing their young, impressionable audience to buy digital stickers during live streams.


Regarding manipulation, our participants share underlying factors relating to a desire for more self-determination and protection of vulnerable users.
Specifically, participants report uncontrollable app behavior {\footnotesize[P11]} or functionalities that are solely implemented to boost engagement {\footnotesize[P8, P9]}.
Nonetheless, their main concern lies with vulnerable groups that they do not consider to be as aware as themselves, namely children {\footnotesize[P4, P5, P11, P15]}, 
and other vulnerable adult users {\footnotesize[P10, P14, P15]}.





\vspace{-2mm}
{\color{white}\tiny
\subsubsection*{Safety}$ $\\}
\raisebox{-1.25mm}{\includegraphics[width=25mm]{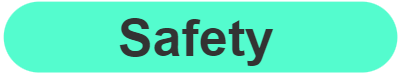}}
Seven participants have active safety concerns, where software presents both a risk physically and mentally.

\paragraph*{Third party threats}
Participants are worried about potential threats from third parties {\footnotesize[P5, P7, P8, P11, P12, P14]}.
P12 believes that online content could lead to violent riots, while P7 feels unsafe in general to speak about certain topics online.
Participants identify further risk figures such as bot accounts, or pedophiles:
\begin{quote}
    \textit{``[...] Pedophiles might text [children], and [the children] may be like, oh, this person loves me, [...] that's a lot of how these kidnapping and sex trafficking schemes work."} -P5 Gen Z Man
\end{quote}



\paragraph*{Health risks} 
Our participants are worried about how software can affect people's mental and physical health {\footnotesize[P1, P4, P12, P15]}.
Mental health concerns often relate to worries about addiction and manipulation.
Social media applications such as TikTok can be seen as a health risk {\footnotesize[P1]}, the usage of which can easily lead to an unhealthy lifestyle. 
They are concerned over harmful targeted content that, e.g., influences people to buy untested, harmful hygiene products{\footnotesize[P4]}, or leads some toward fighting and other violent actions {\footnotesize[P4, 12, P15]}.



Similar to the manipulation concern, our participants are often more concerned about other's safety.
Apart from targeted and manipulative content {\footnotesize[P4, P12, P15]} that may lead people to harm themselves and/or others, participants underlying factors for their safety concerns relate to the pervasive collection of personal data {\footnotesize[P8, P14]}, and the existence of malicious actors, including bots, on the platform {\footnotesize[P5, P8, P11]}.
All underlying factors thus stem, in the case of safety, from other ethical concerns.

\vspace{-2mm}
{\color{white}\tiny
\subsubsection*{Inappropriate content}$ $\\}
\raisebox{-2.75mm}{\includegraphics[width=25mm]{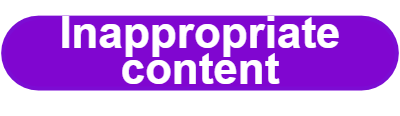}}
Six participants express active concern over inappropriate content.
Participants generally relate this concern to social media.

\paragraph*{Sexual content}
Four participants worry about sexual content {\footnotesize[P5, P7, P16, P17]}. 
They describe how other users show their genitalia {\footnotesize[P5]} and create bot accounts with profile pictures of naked women {\footnotesize[P17]}.
They are concerned that AI is now used to generate pornography {\footnotesize[P16]}.
P7 is concerned that children will participate in adult trends:
\begin{quote}
    \textit{``[...] some sexualized dances, which I think is really wrong."} -P7 Gen Z Woman
\end{quote}
Here, P7 is not only concerned about what children can see, but also about child safety, and who will end up seeing content that children reproduce.

\paragraph*{Negative content}
Some believe that the diffusion of some negative content is unethical due to the current nature of social media platforms {\footnotesize[P9, P12]}.
They explain that discussions about certain topics are guaranteed to turn into fighting in these environments.
P12 lists as examples topics relating to politics, the economy, religion and social standing.

P8 mentions explicitly that some violent content is disturbing. 
She points out that websites such as 4Chan.org are not regulated and spread video content depicting actual torture, murder and rape. 

The underlying factors for concerns about inappropriate content are: that it could create a wrong idea of reality {\footnotesize[P16]}, that it caters to predators {\footnotesize[P7]}, that it is something age-inappropriate {\footnotesize[P5]}, that it is highly disturbing {\footnotesize[P8]}, and that it can lead to riots and online threats {\footnotesize[P12]}.

\subsection{Latent ethical concerns (RQ2)} \label{sec:latent}
We ask participants specifically if they are concerned about addiction, censorship, and discrimination in relation to the software they use.
Figure~\ref{fig:ec_frequency} shows that participants rarely bring these concerns up freely in the unprimed phase.
When specifically asked, 11 participants express concern on addiction, 13 on censorship, and 12 on discrimination.

        
        

\vspace{-2mm}
{\color{white}\tiny
\subsubsection*{Addiction}$ $\\}
\raisebox{-1.25mm}{\includegraphics[width=25mm]{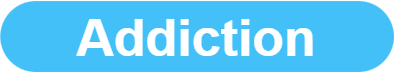}}
Eleven participants demonstrate concerns about addiction, only one {\footnotesize[P19]} needed an example. 
\paragraph*{Social media addiction}
When thinking of addiction, a majority of participants shares concerns about social media addictions.
They specifically name TikTok, Instagram, Twitter, Facebook, and Snapchat in their answers {\footnotesize[P1, P4, P5, P6, P7, P8, P10, P11, P13, P14, P16, P17, P18]}. 
Participants share that they observe others engage in \textit{doom-scrolling}, i.e., looking for ever more stimulating, relatable content {\footnotesize[P1, P6, P7]}.
P14 says that she is scared that users absorb online content "like zombies".

Underlying technical factors of social media addiction are specific elements such as the video reels format {\footnotesize[P7, P14]}, likes and comments {\footnotesize[P1, P14]}, notifications {\footnotesize[P7]}, streaks  {\footnotesize[P8]}, infinite scrolling {\footnotesize[P1, P6, P7]}, the sheer volume of available content {\footnotesize[P4, P14]}, and personalized algorithms {\footnotesize[P17]}. P17 mentions the personalized algorithms as "fun algorithms".
Additionally, participants links social media addiction to human tendencies such as the need for communication {\footnotesize[P6, P17]}, validation {\footnotesize[P4, P8, P16]}, stimulation {\footnotesize[P1, P18]}, FOMO (`Fear of missing out'){\footnotesize[P11]}, and natural curiosity {\footnotesize[P7, P17]}. 

\vspace{-2mm}
{\color{white}\tiny
\subsubsection*{Censorship}$ $\\}
\raisebox{-1.25mm}{\includegraphics[width=25mm]{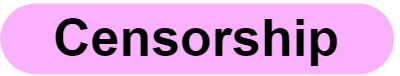}}
A total of 13 participants express concern about censorship in software, seven of which need an example to understand how the concern relates to software.
In relation to inappropriate content, three participants {\footnotesize[P4, P8, P14]} clarify first that some content, e.g., extreme, far right or pedophilia content, should not be allowed on any platform.
Nonetheless, the majority of participants believe it is unethical to censor on social media platforms.

\paragraph*{Loss of freedom}
Without needing an example, participants express their fear of restrictions on the freedom of speech {\footnotesize[P4, P9, P14]}.
P14 hints at \textit{transparency} when she criticises how users are unaware of who is censoring. 
P9 touches on \textit{manipulation} when he mentions how social media platforms can use their terms and conditions, that everyone needs to sign, as a means to restrict freedom of speech. 
He explains that it is unfair to people with a large following to have their posts taken down, because the platform should not subjectively decide what can be shown. 

\paragraph*{Propaganda}
P8 similarly describes how posts of her friends have been censored for being pro-Palestine and how their content views instantaneously fell from 500 to 60.
She believes the accounts were shadow-banned (when content is not fully banned, but demoted in the content streams to reduce its visibility).
P7 expresses a general distrust in media, saying that it is all controlled and that "the truth is not out there on social media".
P3 summarizes the position of the majority of our participants:
\begin{quote}
    \textit{``[...] an app in its own should not already pick a side and then censor the other side. I think it should be a neutral place to be honest."} - P3 Gen Z Woman
\end{quote}

Ultimately, the underlying factors of the censorship concern are that social media content shapes the views of the public and should therefore not be controlled by a single entity {\footnotesize[P2, P3, P16, P17]}, and that censorship restricts freedom of speech {\footnotesize[P4, P5, P9, P14, P15]}, limits the free flow of information {\footnotesize[P7]}, and makes the platform unfair to those that rely on it to make a living {\footnotesize[9]}.

\vspace{-2mm}
{\color{white}\tiny
\subsubsection*{Discrimination}$ $\\}
\raisebox{-1.25mm}{\includegraphics[width=25mm]{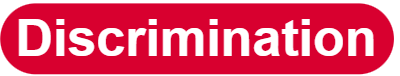}}
Twelve participants show concern about discrimination when primed, nine of which require an example.
As a consequence, the following statements are strongly influenced by our example of a biased automated job application system (see Table~\ref{t:concern_taxonomy}).
Participants unequivocally condemn discrimination, P1 and P14 are appalled that such an algorithm was even conceived.
P7 and P13 believe that discrimination in hiring practices is not going away and that software simply continues the age-old trend.

\paragraph*{Mass-profiling}
Participants that do not need an example take discrimination in its literal sense--to distinguish {\footnotesize[P8, P9]}. 
The fact that users are discriminated based on particular attributes is sufficient to raise concern, as it can be used to skew users perception of reality through targeted content {\footnotesize[P8]}. 
\begin{quote}
    \textit{``
    [...] what the word discriminate means is being able to differentiate people from each other. [...] 
    That in itself is already concerning, because if it can create a specific view of [people, e.g.], this is a man who's 50 years old, who's white. Then it creates an algorithmic procedure of what this person will engage with, and if it sees me, who is a person of color, probably a trans person, then it creates this image of what I would want to see, and that discrimination is already kind of crazy because my reality of my social media is entirely different from somebody else's."} -P8 Gen Z Woman
\end{quote}
P9, sharing the same concern with P8, clarifies:
\begin{quote}
    \textit{``[...] political views, social concerns, social values, or content that are in favor or against some social values, I think those are some things that should not be targeted or discriminated against, right?"} -P9 Gen Z Man
\end{quote}

We gather as underlying factors that our participants essentially see discrimination as a tool for manipulation {\footnotesize[P8, P9]}.
Their worries about privacy and censorship support their view, pointing at the highly personalized content on social media platforms.
With respect to our example, participants believe that personal demographic attributes should not matter in hiring practices, and that it would be unfair {\footnotesize[P1, P2, P3, P4, P12]}.

\subsection{End-user's solutions (RQ3)} \label{sec:solutions}
To gather potential software solutions, we ask our participants what they think needs to change to dispel their ethical concerns.
In their answers, we identify \textit{internal} changes to software, directly affecting the application space, and \textit{external} changes, depending on decisions of various stakeholders such as governments and business management.
We present six suggestions towards more ethical software practices that we identify from participants' perspectives, four of which are internal and two external.



\subsubsection*{Application information clarity \textit{(Internal)}} 
Participants believe software applications need more clarity and additional explanations {\footnotesize[P2, P6, P8, P12, P13, P14]}. 
This requirement addresses the ethical concerns of transparency, privacy, misinformation, and censorship.
They express the need for more user-friendly consent forms that clearly outline what information is being collected and why {\footnotesize[P2, P6, P8]}.
P8 points out that software practitioners need to be serious about user consent to improve software transparency: 
\begin{quote}
    \textit{``[...] in a morally ethic, perfect world you would click yes and they ask, are you sure? And then it would be a box popping up this is what we do with your data. This is what's possible. This is what we know about you."} - P8 Gen Z Woman
\end{quote}
P12 highlights that visible assurances of personal data protection could provide a "guarantee" and give her more confidence in sharing personal data.
Additionally, participants emphasized the need for better communication regarding when and why and by whom content is censored {\footnotesize[P13, P14]}. 
In general, users search explanations for how things come to be, not only in their social media feeds but also in services such as ChatGPT, that must provide sources and explanations for the output it generates [P14].

\subsubsection*{User control and experience (Internal)}
Participants desire greater control over software applications and want a better user experience to address addiction, manipulation, and privacy concerns {\footnotesize[P1, P5, P7, P8, P9, P11, P12, P17]}. 
They advocate for the removal of addictive features such as likes {\footnotesize[P1, P8]}, streaks {\footnotesize[P8]}, comments {\footnotesize[P1, P8]}, video reels {\footnotesize[P7]}, and intrusive notifications {\footnotesize[P7]}, along with enhanced anonymity {\footnotesize[P1]}, to mitigate the addiction to social media applications. 
P7 thinks the novel video reels feature added recently to major social media applications is and "evil" idea---
intentionally designed to be addictive. 
She comments:
\begin{quote}
    \textit{``[...] Because it's not screened, it's very addictive, it decreases attention span, possibly like your eyesight gets worse because you keep looking at your phone like this [...].
    So I think I personally would prefer if it just wasn't there in the first place."} - P7 Gen Z Woman
\end{quote}
Participants further demand more control over personal data, including a universal option to turn off their microphone {\footnotesize[P9]} and more user-friendly cookies {\footnotesize[P11]}.

\subsubsection*{Content control and filtering (Internal)}
Participants {\footnotesize[P3, P8, P9, P11, P12, P17]} advocate for content control and filtering on social media to address censorship, inappropriate content, manipulation, discrimination, and harmful advertising concerns, emphasizing platform neutrality and a balanced 
media diet {\footnotesize[P3, P8, P9, P17]}. 
Participants suggest that recommendation algorithms should  filter less and also serve diverging content 
to avoid engagement-driven content biases {\footnotesize[P8, P9, P11, P17]}.
P17 believes that apps should lead users to have less extreme views on sensitive topics such as medicine and politics: 
\begin{quote}
    \textit{``[...] The app should actually ensure that there are also opposing videos against, so that its possible to not only choose one side, but simply continue to hear both sides."} - P17 Millennial Woman
\end{quote}
Giving users better control over content filtering could reduce worries over censorship and manipulation concerns.
P8 adds that features to report harmful content need to be more responsive.
With respect to operating systems, P11 and P12 suggest that they are not suited for some content, such as advertisements {\footnotesize[P11]} or negative content {\footnotesize[P12]}. 
P11 thinks that companies are misusing user interface elements meant for warnings and important system information.

\subsubsection*{Safety and user protection (Internal)}
Participants propose restrictions in software applications for mental and physical safety concerns.
Social media platforms should implement warnings {\footnotesize[P14]} and time limits {\footnotesize[P1, P5, P12, P17]} to prevent excessive screen time.
Children should be protected by restricting their access to certain content {\footnotesize[P5, P7, P17]}.

\subsubsection*{Humane approach (External)}
Participants advocate for more humane approaches in software.
They believe it is wrong to discriminate users by their demographic attributes {\footnotesize[P2, P3, P12, P15]}.
Participants perceive our example of a biased job application algorithm as unacceptable, others mention the \textit{Toeslagaffaire}~\cite{theguardianDutchGovernment} in the Netherlands as an example of gross software-related misconduct.
Participants suggest that a human element cannot be skipped when applying software to deal with large amounts of data, e.g., when reviewing flagged content or accounts {\footnotesize[P2, P15]}.
P15 suggests that reviewers could apply \textit{``a more personal approach"}, potentially by directly communicating with the accused parties.

\subsubsection*{Government control and regulation (External)}
The majority of participants call for increased government control and regulation of applications {\footnotesize[P2, P4, P8, P9, P10, P11, P12, P13, P14, P16, P17, P19]}. 
They emphasize that the government needs to protect personal data {\footnotesize[P10, P11, P14]}, e.g., by hosting it on European servers:
\begin{quote}
    \textit{``[...] The usage of that data can be supervised by independent European Government Inspectors or something like that, so that it's sure it's not going anywhere and that they delete the data [...]."} -P11 Gen X Woman
\end{quote}
Additionally, they called for laws and regulatory bodies to oversee AI and algorithm implementation {\footnotesize[P9, P14, P16, P17]}. 
Participants further suggest to restrict who can use certain applications {\footnotesize[P9]} and monitor inappropriate content in applications {\footnotesize[P4, P8, P12]}.

\section{Discussion}

Our results show that:
(RQ1) participants actively worry about ethical issues in software, mainly about privacy, transparency, manipulation, safety and inappropriate content,
(RQ2) participants latently worry about software addiction, and also about censorship and discrimination---when explained how they relate to software; and
(RQ3) participants believe their worries could be alleviated through better clarity and control in applications, through more humane approaches to software and increased government regulation.
We revisit each research question as follows:

\textit{RQ1 on active ethical concerns: }
Figure~\ref{f:ec_map} shows that participants' worries often relate back to safety and manipulation concerns. We discuss this interweaving next.

Our findings on manipulation signal that users want to feel in control of their own digital life. 
They want to understand how their data is processed within applications they often interact with because this knowledge could give them the ability to better assess the situation and predict future outcomes.
The absence of understanding leads to worries over bad intentions behind opaque practices.
Participants feel actively surveilled, and are frequently reminded of it by omnipresent targeted advertising.
Another constant reminder of powerlessness are consent forms; extremely detailed documents that take too long to read because they are extensive and written in unfamiliar legal language, often believed to hide dishonest intentions. Previous theoretical research mirrors this finding, critiques of current digital consent practices point out their use maintains opacity through an extreme power imbalance~\cite{carmi2021feminist, carrigan2021revolution}.
Participants also identify dark design patterns' manipulative power as discouraging, especially within social media. 


While participants use technology full of risks, they are protective of those they consider vulnerable and less aware.
They report seeing children consume harmful content and fear them coming in contact with abusers.
They name harmful effects such as social isolation, a reduced attention span and a skewed world view due to experiencing it mainly through social media.
These participants are rightfully concerned: previous research has revealed that anonymous interaction with harmful online content is frequent among adolescents and might be related to higher depressive levels and initial suicide ideation~\cite{pater2017defining, patchin2017digital}.

As a result, participants point out the need for more protection and guidance for children online.
Our participants report remembering a time where they felt safe online, being able to browse the web in complete anonymity, perhaps suggesting that some applications could be improved through technological regression. 





\textit{RQ2 on latent ethical concerns: }
Perspectives of latent ethical concerns are frequently tied back to previously mentioned active concerns. When we ask our participants about concerns of addiction they mainly name social media platforms, specifically video reels as culprits.
They describe that such software does not benefit their personal growth, consumes a lot of time and negatively influences emotions.
Multiple participants call for the regulation of addictive software due to its profitable nature. Recent research confirms this suspicion: addictive, sensational content is generated because there is a financial incentive to do so~\cite{munger2022right}. 

Participants view censorship as complex, accepting restrictions on explicit content but expressing concerns over the suppression of political ideologies. By contrast, app review analyses highlight frequent complaints about conservative viewpoint restrictions~\cite{Tjikhoeri2024}. In-person discussions, therefore, revealed less overtly ideological stances, suggesting more moderated perspectives in direct interactions.

When asked about discrimination concerns in software, some participants name the \textit{Toeslagaffaire}~\cite{theguardianDutchGovernment}, a case that attracted widespread media attention after the Dutch government admitted to having wrongfully accused families of childcare benefit fraud with the help of a racially-biased algorithm~\cite{autoriteitpersoonsgegevensMethodsUsed}; indicating that widely reported events heavily influence perceptions of ethical concerns in software. 
The latter has also been recognized in previous work analyzing ethical concerns of intersectional communities expressed on social media~\cite{olson2024crossingmarginsintersectionalusers}
Discrimination is related to manipulation, as participants recognize that personalized experiences could be turned against the user.  

Figure~\ref{f:ec_map} is a supplementary extension to the taxonomy of ethical concerns established in prior work~\cite{olson2023, Tjikhoeri2024}. While the earlier taxonomy categorized ethical concerns into discrete groups, Figure~\ref{f:ec_map} advances this foundation by introducing secondary themes and elucidating connections between the categories. Although our version offers enhanced granularity, it encompasses only a subset of the ethical concerns included in the original taxonomy. Future research should focus on its extension to a broader range of ethical concerns.




\textit{RQ3 on potential solutions for ethical concerns: }
Our participants suggest that practitioners implement software features with high information clarity, especially for consent forms.
This does not necessarily mean that applications contain more explanatory text, since users are already overwhelmed by text volume~\cite{meier2020shorter}.
Alternative solutions to improve user consent emphasize dynamic systems that delegate decision-making to trusted third parties, enhancing fairness and reducing user burden~\cite{nissen2019should}. AI-driven chatbots also improve usability by making consent forms more readable and promoting user agency~\cite{Xiao_2023}. To address privacy and transparency concerns, participants suggest apps prevent echo chambers on sensitive topics like medicine and politics. Practitioners can achieve this by integrating fairness principles into recommendation algorithms to promote diverse perspectives and reducing financial incentives for polarizing content~\cite{munger2022right}.

For managing harmful content, participants advocate for overhauling reporting mechanisms by dedicating more resources to reviewing flagged content, a practice aligned with improving transparency. Providing personalized explanations for content removal decisions and implementing mediation systems with trained psychology professionals could further enhance user trust and safety. Lastly, participants highlight the need for tools to manage excessive screen time. Practitioners should develop usability-focused features that empower users to monitor and control their screen time, fostering autonomy and improving digital well-being while adhering to principles of fairness and user-centric design.
The collection and processing of data should be self-explanatory, and purposes of the collected data should be announced frequently, not only in the beginning when agreeing to conditions of use.
We note that practitioners should not infantilize users, as they are able to grasp complex processes if they are well explained.
On this point, researchers---many also educators---must envision new methods for achieving an understanding of hidden processes in applications, through the applications themselves.

Some software functionalities have such negative perception, that participants can only think of removing them entirely.
We recommend that practitioners allow users to opt out of addictive features such as likes, comments, video reels especially, as well as  algorithms designed to artificially increase user interaction.
One participant suggests that practitioners should be directly exposed to user feedback about ethical concerns to foster a sense of responsibility.


Computer technology has advanced significantly in recent years, yet the rapid pace of these development outpaces legislation, leaving it imperfect and essential for addressing the ethical concerns presented in this work. 
We believe that the introduction of the AI Act (AIA)~\cite{european-parliament-no-date} in May 2024 addresses many concerns of our participants.
We also see GDPR as a good development, but participants are correct to point out that it does not cover all their privacy concerns. For example, GDPR addresses both consent forms and cookies~\cite{edpb2020consent}, and yet users still have significant complaints regarding transparency regarding these features.
An interesting approach for addressing many of these is the browser add-on ToS;DR~\cite{tosdr}, which assigns privacy scores to online services, not dissimilar to the risk scores given to AI products through the AIA. 
For example, the government could for example treat addictive software similarly to addictive substances and require warnings within the user interface. 

The Calm Tech Certification~\cite{calmtechcertification} encourages software design that reduces intrusive, manipulative, and addictive features by promoting minimalism, user control, and non-disruptive functionality. Such certifications could incentivize ethical practices by tying them to regulatory benefits or enhanced public trust.

The ACM~\cite{acm_code} and IEEE code of ethics~\cite{ieee_code} establish abstract, universal principles (e.g., be honest, avoid harm), which are rooted in deontological, utilitarian Western philosophical traditions. This ideology, while useful for broad applicability and consistency, excludes other ethical traditions, like Black Feminism, Care ethics and Indigenous philosophies which emphasize relational, situated, and community-oriented ethics~\cite{birhane2022forgotten}. Our research does not intend to establish principles suitable to software applications broadly as it is not representative of any population. However, we encourage the augmentation of ACM and IEEE codes of ethics's professional responsibilities to include the philosophical stance \textit{epistemic humility}, a ``sensitivity to difference" that urges practitioners to critically examine their own assumptions and positionality, rather than presuming to understand the needs and lived experiences of others~\cite{hollanek2024ethico, ansari2019decolonizing}. Our study’s in-depth findings provide a valuable starting point for practitioners seeking to engage with and operationalize epistemic humility in their work.
Prior research has shown that ethical concerns of users and practitioners often diverge~\cite{jakesch2022different}, making such humility essential for bridging this gap.

\section{Limitations}

This study adopts an in-depth interpretivist approach, emphasizing deep, case-oriented analysis over representativeness. Previous research shows that a sample size of 12–30 participants per country is recommended to perform a profound, case-oriented analysis~\cite{boddy2016sample}, which is the `\textit{raison-d'etre} of qualitative inquiry'~\cite{sandelowski1995sample}. Our sample, all residents of the Netherlands but from diverse cultural backgrounds, reflects this intent. We employed snowball and maximum variation sampling to ensure diversity across age and gender, avoiding restrictions to any specific user group~\cite{suri2011purposeful}. This approach reflects our research goal of identifying common ethical concerns that transcend individual demographics or usage contexts.
The results should not be used to argue about prevalence of one concern over the other, but to begin to understand relationships between distinct perspectives within the local context.

While we have no prior relation to any of the participants, we expect social desirability biases~\cite{nederhof-1985}. We counteracted these by sharing the research goal ahead of time~\cite{bergen-2019}.



Despite peer-reviewing and pilot testing, we find minor limitations in our interview process.
The definition of the discrimination concern allows for unintended interpretations, e.g., equating targeted advertising to discrimination because it differentiates between people.
Participants also assume from the given example that the racial bias was introduced with malicious intent.
In future work we must clarify that discrimination refers to the systemic, unfair treatment of marginalized communities.
We should further consider using more examples, or none, to not prime participants towards a single topic.
 
While we took precautions to reduce potential researcher biases, we are aware that the backgrounds of our researchers do not cover all participant backgrounds, i.e., middle eastern countries, creating a risk of blind spots.
Nonetheless, we minimize the influence of our subjective experiences, through code discussions between the two first authors, the two discussion rounds with all authors and further meetings of three authors to discuss the work.
We describe in detail how we ensure the validity of our coding process in Section~\ref{s:data_analysis} \textit{Data analysis}.

\section{Conclusion}

In 19 semi-structured interviews with individuals across various generations and backgrounds, we uncover a complex mapping of concerns that mostly reveals worries of a social nature.
Our participants care about those closest to them, and the younger generations.
They detail how software can lead to addictions, to safety risks for physical and mental health, and to the deterioration of social skills.
Our participants are sometimes not as worried about themselves, as they are about society as a whole, sometimes even blaming themselves for being addicted to software, or stepping into other digital traps.
Many see the threat of total surveillance in every targeted advertisement, imagining how companies can manipulate people by feeding users what to believe, while censoring the opposition.
Users expect software practices to be humane, and suggest change with respect to information clarity, control over applications and protective measures for vulnerable individuals. 
Government action through legislature, such as the AI Act and GDPR, is commonly suggested as a solution in cases where software practices do not change for the better.
It is therefore crucial for researchers and practitioners to delve deeper into ethical concerns about software, ensuring that ethical considerations drive technological innovation by taking ownership of social responsibilities. 
The path forward demands continuous investigation and proactive measures to align software practices with evolving user expectations and values.

\bibliographystyle{IEEEtran}
\bibliography{references}

\end{document}